\documentclass[aps,pra,twocolumn,citeautoscript]{revtex4-1}%
\pdfoutput=1
\usepackage{amsfonts}
\usepackage{amsmath}
\usepackage{amssymb}
\usepackage{graphicx}%
\usepackage{subfigure}

\begin{document}
\preprint{ }
\title{Atomic Fermi gas at the unitary limit by quantum Monte Carlo methods: Effects of
the interaction range}

\author{Xin Li,$^1$ Jind\ifmmode \check{r}\else \v{r}\fi{}ich Koloren\ifmmode \check{c}\else \v{c}\fi{},$^{1,2}$ and Lubos Mitas$^1$}
\affiliation{$^1$Department of Physics, North Carolina State University, Raleigh, NC 27695, USA \linebreak $^2$
I. Institut f\"ur Theoretische Physik, Universit\"at Hamburg,  Jungiusstra\ss e 9, 20355 Hamburg, Germany }

\keywords{unitary limit, fixed-node diffusion Monte Carlo, released-node diffusion Monte Carlo}

\date{\today}

\begin{abstract}
We calculate the ground-state properties
of unpolarized two-component Fermi gas by
the diffusion quantum Monte Carlo (DMC)
methods. Using an extrapolation to the zero effective range of the
attractive two-particle
interaction, we find $E/E_{\rm free}$ to be
0.212(2), 0.407(2), 0.409(3) and 0.398(3) for 4, 14, 38 and 66 atoms,
respectively. Our results indicate that the dependence of the total energy on the effective range
is sizable and the extrapolation is therefore quite important.
In order to test the quality of nodal surfaces and to estimate the impact
of the fixed-node approximation
we perform released-node DMC calculations for 4 and 14 atoms. Analysis
of the released-node and the fixed-node results suggests
that the main sources of the fixed-node errors are long-range correlations which
are difficult to sample in the released-node approaches due to the fast growth of
the bosonic noise. Besides energies,
we evaluate the two-body density matrix and the condensate fraction. We
find that the condensate fraction for the 66 atom system converges
to 0.56(1) after the extrapolation to the zero interaction range.

\end{abstract}
\maketitle

\section{Introduction}
In recent years, the
homogeneous Fermi gas with attractive interactions has been studied extensively both theoretically and experimentally
due to the success in cooling atoms into ultracold dilute condensates
\cite{Ketterle02, Ketterle08, Giorgini08}. By
tuning the interaction strength through the
Feshbach resonance,\cite{Fano61, Feshbach62, Ketterle98, Verhaar98} the system can cross from the
Bardeen-Cooper-Schrieffer (BCS) superfluid phase, where the s-wave scattering length $a_s$ is
negative, to the Bose-Einstein condensate (BEC),
where $a_s$ is positive. Since there is no symmetry change of the
quantum state involved,
the system exhibits the well-known BCS-BEC crossover.

In the special case corresponding to the diverging scattering length,
$a_s \to \infty$,  the system is in a strongly interacting regime called the unitary limit.
In this regime the interparticle spacing $r_s$ is the
only relevant scale, and the rest of the quantities are universal and system independent.
The total energy of this system can be conveniently written
as $E= \xi E_{\rm free}$, where $E_{\rm free}$ is the energy of the non-interacting
atomic gas and $\xi$ is a system independent parameter. Experimental measurements of $\xi$ have
been performed using $^6$Li and $^{40}$K atoms by investigating the
expansion rate of the atomic cloud and the sound propagation in it\cite{Bartenstein04,
Bourdel04, Thomas05, Stewart06, Thomas07}.
Simultaneously, a number of theoretical and numerical estimations of $\xi$ have been reported,
including diffusion Monte Carlo (DMC)
\cite{Carlson03,Carlson04,Astrak04,Astrak05,Pandharipande05,Needs10} as well as path
integral Monte Carlo, lattice simulations and analytical methods\cite{Lee06, Lee08, Trivedi07,
 Burovski06, Gurarie07, Bulgac06, Bulgac08, Abe09a, Abe09b, Needs10}. The resulting
estimates fall between $\approx$ 0.25--0.45
showing that the actual value has not been settled yet and is still of significant
interest due to the universal nature of the unitary limit.

One of the most interesting properties of the unitary gas is the robust
presence of the pairing condensate which involves a large fraction of the
system.
The study of pairing effects is thus much more straightforward
than, say,  in superconducting materials, where only a sliver of the fermions around
the Fermi level forms the condensate and the attractive interaction is much more complicated.
The quantum Monte Carlo (QMC)
methods have the advantage that the condensate can be detected directly,
by evaluating the off-diagonal two-particle density matrix and by monitoring
its behavior at large distances
\cite{Astrak05, Needs10}.

The goal of our study is twofold. First, any actual simulation involves
only a finite system, and the quantities relevant for the
thermodynamic limit have to be obtained from appropriate
extrapolations. The unitary
system is not trivial in this respect, since pairing with infinite
scattering length is described by a function with a slow fall-off
at large distances. It is necessary to analyze the finite-size
scaling of the quantities of interest and to test whether the actual limit of
infinite dilution, or, equivalently, of point-like character of the
interaction, has indeed been reached. Second, the impact of
the fixed-node approximation in the quantum Monte Carlo method
is not very well understood for this system since there is nothing
to compare with: so far the fixed-node formulation
of the QMC methods appears to be the only approach
that is able to provide an upper bound for the total energy.
This has motivated us to probe the accuracy of the nodes
by released-node QMC simulations and by improvements in the variational
flexibility of the employed wave functions.

We have carried out calculations
of the ground-state properties of the dilute unitary Fermi gas by the
fixed-node DMC (FN-DMC)\cite{Mitas01} method for 4, 14, 38 and 66
atoms.  By using a more versatile
construction of the pair orbital in the BCS wave function
and by extrapolating the effective range of the two-particle
interactions $R_{\rm eff}$ to zero, we were able to obtain lower $\xi$
than, for example, the previously reported DMC calculations in Ref. \cite{Carlson03}
based on BCS trial functions.
These results suggest that
the extrapolation of $R_{\rm eff}$ is important,
especially for smaller systems.
In order to test the quality
of the nodal surface of the BCS wave function,
we have performed released-node DMC (RN-DMC)\cite{Ceperley80,Ceperley84} calculations
for 4 and 14 atoms. This procedure has been carried out starting from two types of
nodal constraints: from the BCS nodes and from the Hartree-Fock (HF) nodes.
Our RN-DMC results indicate that the nodal corrections
are driven mainly by long-range correlations which are difficult to sample
in the released-node framework due to the rapid growth of the bosonic noise.
We have calculated also the two-body density matrix and the condensate fraction
for the 66 atom system, and we
have estimated the corrections from the effective-range extrapolation on these quantities.


\section{method}
\subsection{Hamiltonian}
We consider a two-component Fermi gas with Hamiltonian
\begin{equation}
H=-\frac{1}{2}\sum_{i=1}^{N/2}\nabla_i^2 -
\frac{1}{2}\sum_{i'=1}^{N/2}\nabla_{i'}^2
+ \sum_{i,i'}V(r_{ii'})\,,
\end{equation}
where $N$ is the total number of atoms, $i$ and $i'$ correspond to
the spin-up and spin-down atoms, and $r_{ii'}$ denotes the distance $|\mathbf{r}_i-\mathbf{r}_{i'}|$.
The atoms are located in a cubic box
with the side $L$ and we impose the periodic boundary conditions. The
two-particle potential $V(r_{ii'})$ is taken in the P\"oschl-Teller
form
\begin{equation}
\label{eq:pot_PT}
V(r_{ii'})=-\frac{2\mu^2}{\cosh^2(\mu r_{ii'})}\,,
\end{equation}
whose effective range is $R_{\rm eff}=2/\mu$. The s-wave scattering
length $a_s$ is infinite for all values of $\mu\neq 0$.


\subsection{Trial wave functions}
In the majority of our calculations
we employ trial wave functions of the BCS
form multiplied with the Jastrow factor (BCS-Jastrow) as given by
\begin{equation}
\Psi_T(\textbf{R})=\Psi_{BCS}(\textbf{R})e^{J(\textbf{R})}\,,
\end{equation}
where
\begin{equation}
\Psi_{BCS}(\textbf{R})= \mathcal{A}\biggl[\prod_{i,i'=1}^{N/2}\phi(\mathbf{r}_i,\mathbf{r}_{i'})\biggr]
=\mathop{\rm det} [\phi(\mathbf{r}_i,\mathbf{r}_{i'})]\,.
\end{equation}
Here $\mathcal{A}$ represents the antisymmetrization operator and
$\phi(\mathbf{r}_i,\mathbf{r}_{i'})$ is the pair orbital. The vector
$\mathbf{R}$ encompasses all atomic coordinates $\mathbf{r}_i$ and
$\mathbf{r}_{i'}$. Additionally,
we have carried out a subset of calculations also
with the Hartree-Fock-Jastrow (HF-Jastrow) trial
functions, in which $\Psi_{BCS}$ is replaced with a product
of two Slater determinants of one-particle orbitals (simple plane
waves). The HF-Jastrow wave
function reads as
\begin{equation}
\Psi_{SJ}(\textbf{R})= \mathop{\rm det}[\varphi_a(\mathbf{r}_i)] \mathop{\rm det}[\varphi_a(\mathbf{r}_{i'})] e^{J(\textbf{R})}.
\end{equation}

The pair orbital $\phi(\mathbf{r}_i,\mathbf{r}_{i'})$ in $\Psi_{BCS}$ is written as a linear combination of
Gaussian functions
\begin{multline}
\phi (\mathbf{r}_i,\mathbf{r}_{i'})=\sum_{l,m,n=-1}^{1}\sum_{k} d_{k}
e^{-\alpha_{k}(x_i-x_i'+lL)^2}\\ \times e^{-\alpha_{k}(y_i-y_i'+mL)^2}e^{-\alpha_{k}(z_i-z_i'+nL)^2},
\end{multline}
where $d_{k}$ are expansion coefficients, and $\mathbf{r}_i=(x_i,y_i,z_i)$
and $\mathbf{r}_{i'}=(x_{i'},y_{i'},z_{i'})$ are coordinates of $i$ and $i'$ atoms inside
the simulation box of linear size $L$.
We choose sufficiently large exponents $\alpha_{k}$
so that only the first neighbor shell of periodic images
contributes to the sum, that is, the Gaussian functions are negligible at distances
larger than $3L/2$. The pair orbital is smooth with zero derivative at
the boundary of the simulation cell.
The Jastrow factor $J(\textbf{R})$ is constructed in a similar way as the pair orbital $\phi (\mathbf{r}_i,\mathbf{r}_{i'})$
for both different spin atoms and same spin atoms.

A typical trial wave function includes around 30 to 40 variational parameters that
are optimized by minimizing a linear combination of the total energy and its variance\cite{Umrigar05}.
Although the Jastrow factor does not change
the nodal surface, accurate description of the pair correlations
makes the variational optimization much more efficient and robust.
When the effective range of the potential approaches zero,
more Gaussian functions with larger exponents $\alpha_{k}$ are included in
the Jastrow factor in order to keep the accuracy of the trial function consistently high.
On the other hand, and somewhat surprisingly, we
find that similar adjustment of the pair orbital with the changing effective
range is relatively minor.

\subsection{Fixed-node and released-node DMC methods}
The DMC method projects out the ground state from a given trial function
$\Psi_T$ by means of an auxiliary evolution in the imaginary time,
$\Phi(\tau)\sim\exp(-\tau H)\Psi_T$.
By introducing importance sampling\cite{Mitas01} with the aid of a guiding function
$\Psi_G$, we can write an integral equation for
$\Phi(\textbf{R},\tau)$ in the form
\begin{multline}
\label{eq:DMC_integral}
\Psi_G(\textbf{R})\Phi(\textbf{R},\tau+\Delta\tau)= \\
\int d\textbf{R}' \frac{\Psi_G(\textbf{R})}{\Psi_G(\textbf{R}')} G(\textbf{R},\textbf{R}',\Delta\tau)
                                                    \Psi_G(\textbf{R}')\Phi(\textbf{R}', \tau)\,.
\end{multline}
For small $\Delta\tau$, the propagator $G(\textbf{R},\textbf{R}',\Delta\tau)$ can
be approximated using the Trotter-Suzuki formula as
\begin{multline}
\label{eq:Trotter}
\frac{\Psi_G(\textbf{R})}{\Psi_G(\textbf{R}')} G(\textbf{R},\textbf{R}',\Delta\tau)
\approx G_0(\textbf{R}, \textbf{R}'+\Delta\tau\mathbf{v}(\mathbf{R}'),\Delta\tau) \\
 \times e^{-\Delta\tau[E_L(\textbf{R})+E_L(\textbf{R}')-2E_T]/2} ,
\end{multline}
where $\mathbf{v}(\mathbf{R}')\equiv\nabla\ln|\Psi_G(\textbf{R}')|$ and
$G_0(\textbf{R},\textbf{R}',\Delta\tau)$ is the Green's function for
non-interacting atoms that takes the form of the diffusion kernel. The
so-called local energy $E_L$ is given by
\begin{equation}
E_L(\textbf{R})=\frac{H\Psi_G(\textbf{R})}{\Psi_G(\textbf{R})}\,.
\end{equation}

The product $\Psi_G\Phi$ is represented by a set of samples (also
referred to as walkers) and this ensemble is evolved with the aid of a
stochastic process simulating Eqs.~\eqref{eq:DMC_integral}
and~\eqref{eq:Trotter}. In the fixed-node method we set $\Psi_G(\textbf{R})=\Psi_T(\textbf{R})$ and the fixed-node
condition is imposed by enforcing the sampling points to obey
\begin{equation}
\Psi_G(\textbf{R})\Phi(\textbf{R},\tau) \geq 0
\end{equation}
at all times. In the limit of long $\tau$ the solution
converges towards the lowest-energy state consistent with the boundary
conditions given by the fixed nodes.

\begin{figure}
\begin{center}
\includegraphics[height=3.00in,width=3.25in]{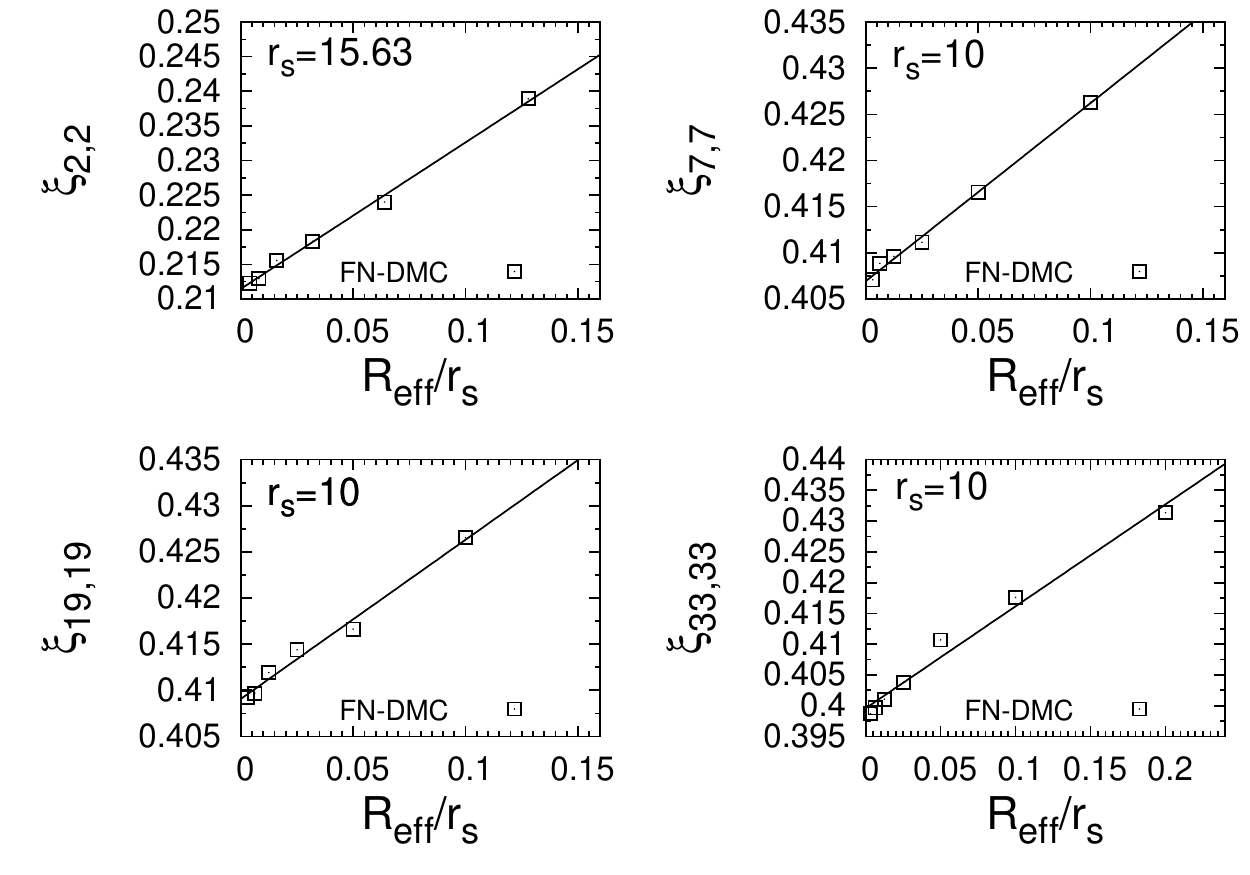}
\end{center}
\caption{The fixed-node energy for unpolarized unitary Fermi gas as a function
of
the interaction range $R_{\rm eff}/r_s$ with linear extrapolation to $R_{\rm eff}/r_s=0$.
The system sizes are 4, 14, 38
and 66 atoms from the top left to the bottom right.
The statistical error bars are smaller than the symbol size.}
\label{tot_extr}%
\end{figure}

In the RN-DMC method the guiding function has bosonic symmetry and
its square should be close to the square of the fermionic ground state.
We have used guiding functions in the form\cite{Ceperley84, Casulleras00}
\begin{equation}
\label{eq:RN_quide}
\Psi_G(\textbf{R})= \sqrt{\Psi_T^2(\textbf{R})+\alpha \left\langle \Psi_{T}^{2} \right\rangle}\,,
\end{equation}
where $\left\langle \Psi_{T}^{2}\right\rangle$ is the average value of $\Psi_{T}^2(\textbf{R}_0)$ over all configurations, and
$\textbf{R}_0$ are the walker positions right after the nodal
release.
The tunable parameter $\alpha$ controls the rate of walkers passing through
the nodal region.
The guiding function is non-negative everywhere and therefore
the stochastic process propagates
a mix of bosonic and fermionic states. The fermionic component
is filtered out
by reweighting with the factor
$\Psi_T/\Psi_G$ so that the fermionic-state
energy is given by
\begin{equation}
\langle\Phi_0|H|\Psi_T\rangle
=\frac{ \int d\textbf{R}\Phi_{0}(\textbf{R})\Psi_G(\textbf{R})\frac{\Psi_T(\textbf{R})}{\Psi_G(\textbf{R})}\frac{H\Psi_T(\textbf{R})}{\Psi_T(\textbf{R})} }
{ \int d\textbf{R}\Phi_{0}(\textbf{R})
\Psi_G(\textbf{R})\frac{\Psi_T(\textbf{R})}{\Psi_G(\textbf{R})} } \,,
\end{equation}
where $\Phi_0(\textbf{R})$ denotes the exact fermionic ground state.

Since this method is exponentially demanding both in the projection time
and in the number of atoms, it is important
to choose $\alpha$ so that the statistical information is recovered as quickly as
possible. If $\alpha$ is too large the fluctuations from the poor importance sampling
overwhelm any useful signal very rapidly. On the other hand, a too small value
can bias the results. Since it is difficult to reach reliable error bars
in this type of calculations, we have used the method mostly to identify the onset
and the amplitude of the energy decrease during the projection
period when the stochastic noise was acceptably small.

In the RN-DMC process, we can also pick up  the statistical signal from the walkers
that have never crossed the nodal surface, and in essence this provides
the FN-DMC estimator.
By monitoring these paths as well, we can assess
the consistency of the estimators and somewhat better
tune the parameter $\alpha$ for providing better RN-DMC signal.


\begin{figure*}
[htb]
\includegraphics[
height=4in,
width=6.2500in
]%
{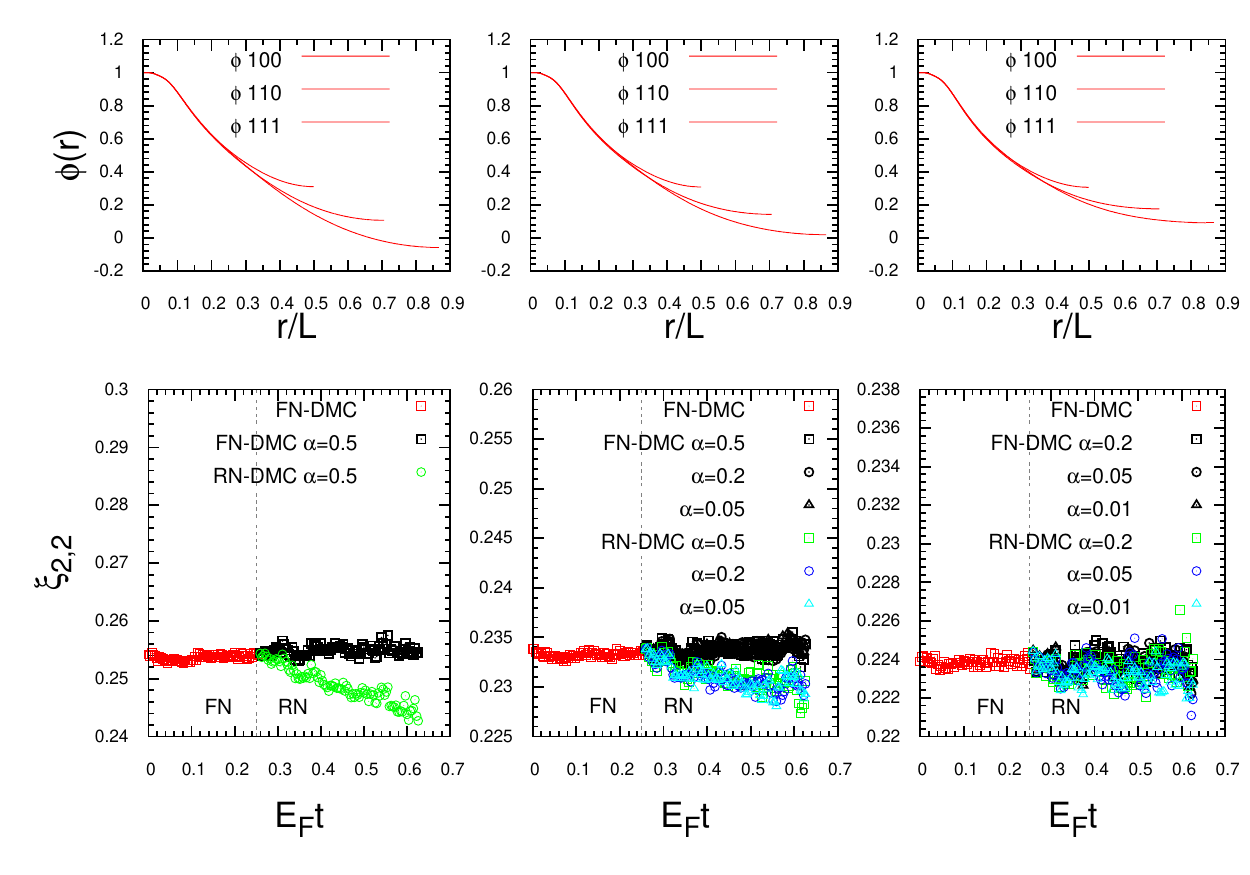}%
\caption{The pair orbitals and FN-DMC and RN-DMC energies of the 4-atom unitary system
with $R_{\rm eff}/r_s=0.06397$.
The upper row shows
the pair orbitals with the lowest (left), intermediate (middle) and optimal
(right) accuracy with regard to the variational optimization.
The lower row shows the corresponding DMC energies
as functions of the projection time starting from the variational estimate.
Note that the resolution of the left and right
panels differs by an order of magnitude. The vertical dotted lines
indicate the instant of the nodal release.
}
\label{4p_wn}%
\end{figure*}

\section{Results}
For benchmark purposes we first
calculate perhaps the smallest nontrivial system---four atoms.
Our result is shown in Fig.~\ref{tot_extr}, upper left panel.
There is approximately $10\%$ energy drop when $R_{\rm eff}/r_s$ is
reduced from $0.1279$ to $0.003998$.
We extrapolate $R_{\rm eff}$
to zero using a linear fit and obtain $\xi_{2,2}=0.212(2)$. Here
and in the rest of the paper, the denominator $E_{\rm free}$ in the ratio
$\xi=E/E_{\rm free}$ is evaluated in the same finite volume subject to
the same boundary conditions as the nominator $E$.

The ground-state energy of this small and relatively simple system was
obtained also by two other numerical methods using a lattice
formulation of the unitary Fermi gas model: the iterative Lanczos
diagonalization and the auxiliary-field projection Monte Carlo
method\cite{Lee11}. The agreement between our fixed-node DMC results
and the outcome of these entirely different methods strongly suggests
that our range-extrapolated total energy of the four-atom system is
very accurate. This conclusion is further corroborated by our
released-node DMC results discussed below.


Our calculations with 14, 38 and 66 atoms are carried out analogously to
the 4-atom case,
and the extrapolated values of $\xi$ are 0.407(2), 0.409(3) and 0.398(3),
respectively, as
plotted in Fig.~\ref{tot_extr}.
In these calculations, the smallest effective range is $R_{\rm eff}/r_s=0.003125$.
It should be noted that in the previous DMC calculations,
the range of $R_{\rm eff}/r_s$ was between 0.1 to 0.2, and within this range our
DMC results agree well with the previously obtained values
\cite{Carlson04,Astrak04,Needs10}.
Reduction of $R_{\rm eff}$ decreases the energy in all cases
although the decrease per atom is smaller in larger systems.

To test the quality of the nodal surfaces, we have carried out
released-node calculations for 4-atom and 14-atom systems.
The RN-DMC calculations for 4 atoms were done with $R_{\rm eff}/r_s=0.06397$.  In a typical released-node
run the number of walkers was about two million so that the error bars were initially
very small.
In Fig.~\ref{4p_wn}, the upper row shows the pair orbital along three
distinct directions (100, 110 and 111)
of the interparticle distance vector ${\bf r}_i-{\bf r}_j$.
The lower row shows the FN-DMC and RN-DMC energies as they evolve with the projection
time. The plots show convergence of the FN-DMC energy followed by the nodal release.
This is accomplished by switching
the guiding wave function from $\Psi_T(\textbf{R})$ to the bosonic
function $\Psi_G(\textbf{R})$ defined in Eq.~\eqref{eq:RN_quide}.

The released-node signal reflects the quality of the nodal surface of
the trial wave function employed in the FN-DMC simulation.
We have tested wave functions with intentionally
varied accuracy by employing suboptimal pair orbitals.
The plot of the energy evolution in the left panel of Fig.~\ref{4p_wn} shows
a clear and pronounced drop after the nodal release.
As the quality of the
pair orbital improves,
this drop shrinks. For the fully optimized BCS-Jastrow
wave function (the right panel in Fig.~\ref{4p_wn})
the energy is reduced by less than 0.002 within the longest projection time we have tried.
This fact as well as comparison with other methods\cite{Lee11}
indicate that our BCS wave functions are very accurate in this small
system and that the fixed-node error is marginal.

We observe an unexpectedly high sensitivity of the nodal quality to the details of the
pair orbital at large distances. This suggests an explanation for the
relatively slow convergence of the released-node energy: the
long-range tails of the pair orbital affect the nodal hypersurfaces,
although their contribution to the total energy is relatively small.
One can further deduce that this makes
the released-node method quite challenging to apply since
it requires sampling of long distances and the corresponding
correlations. This is, however, difficult to achieve because the
diffusive motion of walkers is slow, proportional to $t^{1/2}$, while
the growth of the noise is fast, proportional to $\exp(\Delta_{BF}t)$ where
$\Delta_{BF}$ is the difference between the bosonic and fermionic
ground-state energies.


The time step $\Delta\tau$ was set to $4\times10^{-5}r_s^2$ in all runs and
we have verified that the time-step bias of the RN-DMC results is negligible.
The converged RN-DMC energy should not depend on the parameter
$\alpha$  in the bosonic guiding function
$\Psi_G$ provided one would be able to evolve the stochastic process
with the error bars under control until
the full convergence. In the same time, the parameter $\alpha$ crucially affects the growth
of the fluctuations with the projection time
as illustrated in Fig.~\ref{4p_wn}.

\begin{figure}
\begin{center}
\includegraphics[
height=2in,
width=2.800in
]%
{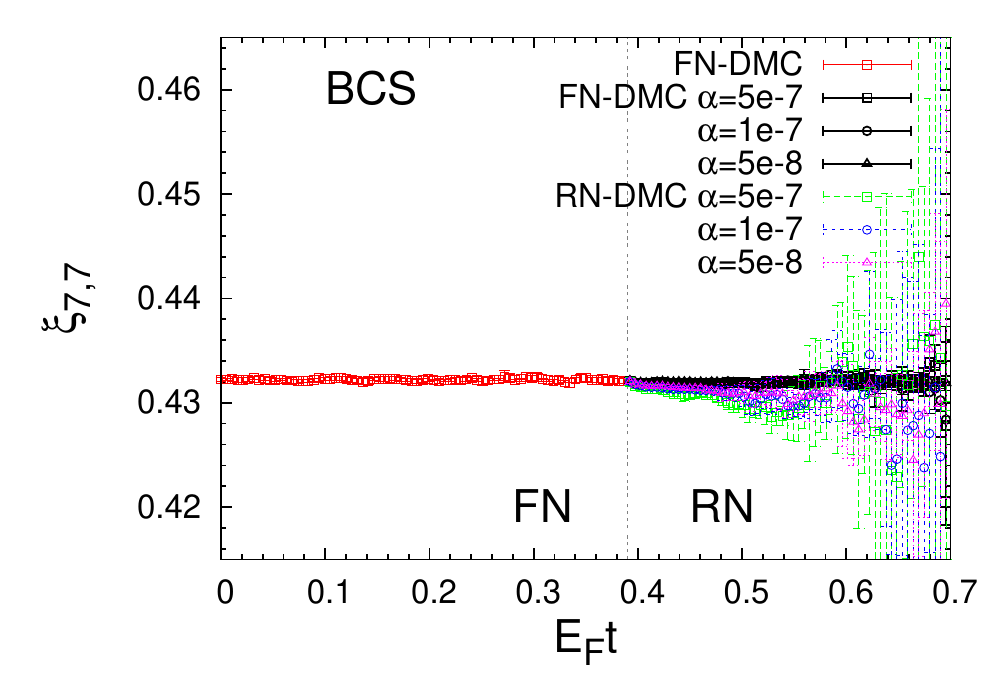}%
\caption{Evolution of the DMC energies for
the 14-atom system with the best optimized BCS-Jastrow wave function.
The runs are for $R_{\rm eff}/r_s=0.2$.
No statistically significant energy drop is observed after the nodal
release that is indicated with the vertical dotted line.}
\label{14p_rn}%
\end{center}
\end{figure}

The RN-DMC energy for 14 atoms with $R_{\rm eff}/r_s=0.2$ is shown in Fig.~\ref{14p_rn}.  The error
bars are estimated from eight independent runs with two million walkers each.
In the interval of $E_Ft\leq 0.2$ after the nodal release the RN-DMC energy gain appears
to be very small and the error bars preclude to make any statistically
sound estimation for longer projection times.
The rapid loss of resolution is expected since
the difference between the bosonic and fermionic ground states grows with
the number of atoms.
Again, the RN-DMC signal exhibits little
dependence on $\alpha$ we choose.

\begin{figure}
\begin{center}
\includegraphics[
height=2.in,
width=2.800in
]%
{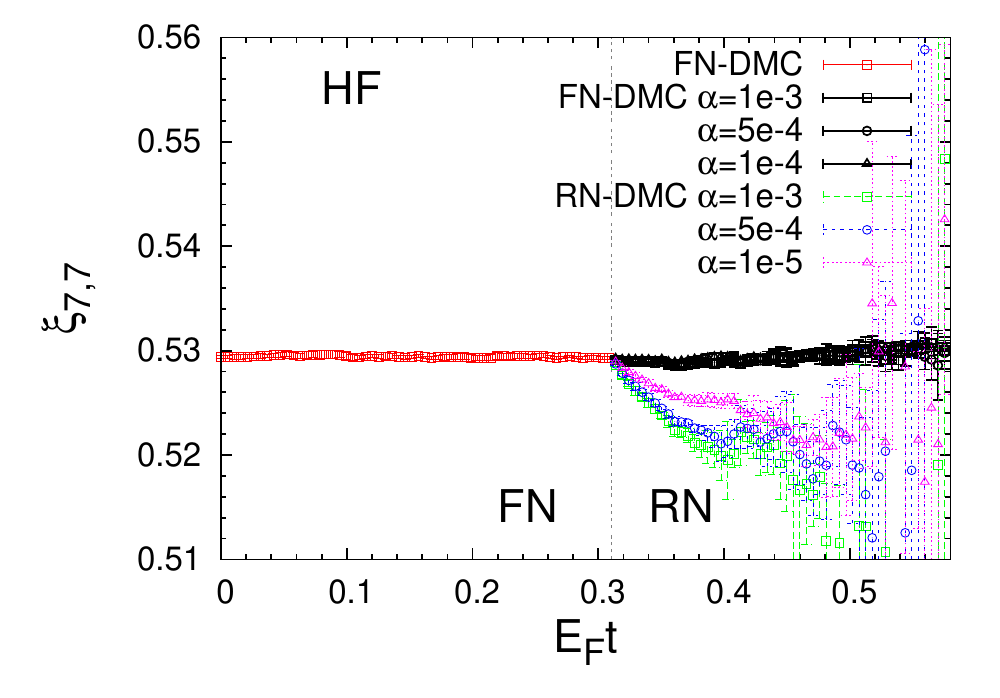}%
\caption{DMC energies for 14 atoms obtained with the Slater-Jastrow wave function.
The runs are for $R_{\rm eff}/r_s=0.2$. The RN-DMC energy drops are significant
when compared
to the RN-DMC signal from the BCS-Jastrow wave function. The parameter $\xi_{7,7}$
drops by $\approx$ 0.015 within $E_Ft\approx0.2$ after the nodal release.}
\label{14p_rn_hf}%
\end{center}
\end{figure}

In order to make a comparison with a case displaying a clear fixed-node bias, we have
carried out RN-DMC runs using the Slater-Jastrow trial wave function, see  Fig.~\ref{14p_rn_hf}.
Since this wave function has the nodal surface of
the non-interacting Fermi gas,
the nodal surface is strongly distorted.
As a result, we see a very pronounced released-node signal.
However,
within the projection time interval of $E_Ft\leq0.2$, the energy drops by
only $\approx 0.015$ for the largest $\alpha$ we tested. This is very small considering
that the true ground-state energy is at least an order of magnitude lower. This again illustrates
the challenges of efficient application of the released-node method, at least for
present cases.
Comparison of the Slater and BCS wave functions shows
a significant effect of pairing as was demonstrated in earlier studies\cite{Carlson03}.

In order to quantify  the pairing effects
we calculate the two-body density matrix which enable us to evaluate the
condensate fraction.
The projected two-body density matrix for spin-up and spin-down atoms is defined as
\begin{equation}
\rho^{(2)}(\textbf{r})=\frac{N^2}{4V^2}
                        \frac{\int d\textbf{R} \Phi(\textbf{R})\Psi_T(\textbf{R}) \frac{\Psi_T(\textbf{r}_1+\textbf{r},\textbf{r}_2+\textbf{r})} {\Psi_T(\textbf{r}_1,\textbf{r}_2)}}
                        {\int d\textbf{R}\Phi(\textbf{R})\Psi_T(\textbf{R})}\,,
\end{equation}
where $N$ is the total number of atoms and $V$ is the volume of the simulation cell.
The density matrices have been calculated for the fixed-node
wave functions and hence they correspond to the mixed estimators
\cite{Mitas01}.
Nevertheless, the mixed-estimator bias is negligible since the
variational Monte Carlo and DMC estimates of $\rho^{(2)}$ coincide within
error bars. This is a further evidence of the high accuracy of our trial
wave functions.

\begin{figure}
[ptb]
\begin{center}
\includegraphics[
height=2in,
width=2.800in
]%
{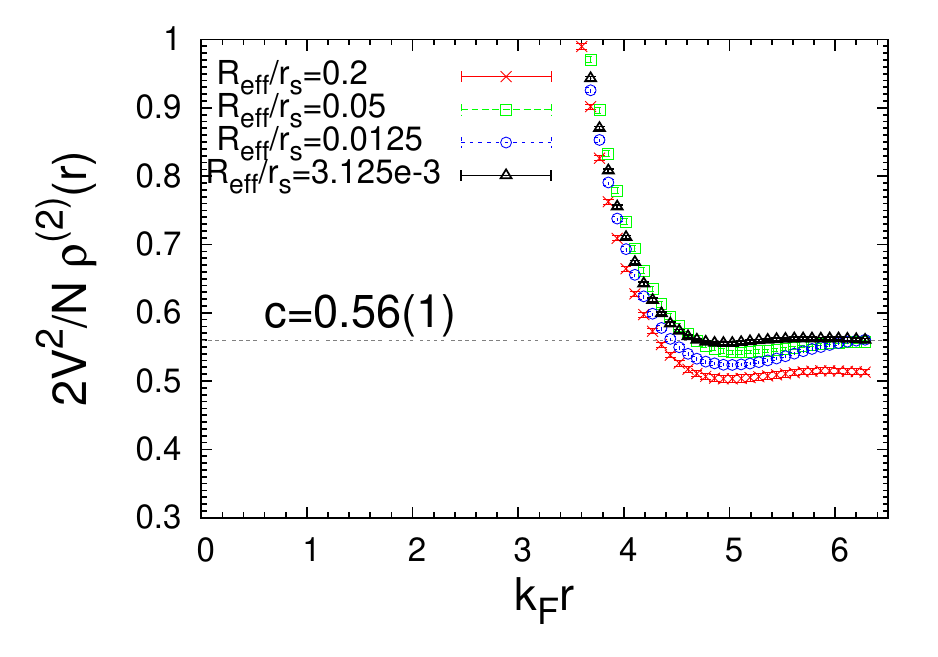}%
\caption{The two-body density matrix for 66 atoms calculated from the
 FN-DMC mixed estimator. The condensate fraction  converges
to 0.56(1) for $R_{\rm eff}/r_s\leq 0.05$.}
\label{66p_rs10_dm}%
\end{center}
\end{figure}

The condensate fraction can be extracted from the two-body density matrix as
\begin{equation}c=\frac{2V^2}{N}\lim_{r \to \infty}\rho^{(2)}(r)\,.\end{equation}
The calculated density matrices are shown in Fig.~\ref{66p_rs10_dm} with
the condensate fraction estimated from the long-range limit.
The condensate fraction saturates
for $R_{\rm eff}\leq 0.5$ at $c=0.56(1)$.
This value is not too far from the results obtained previously \cite{Astrak05,Needs10}.


\begin{figure}
 \centering
 \mbox{
     {\resizebox{1.1in}{!}{\includegraphics{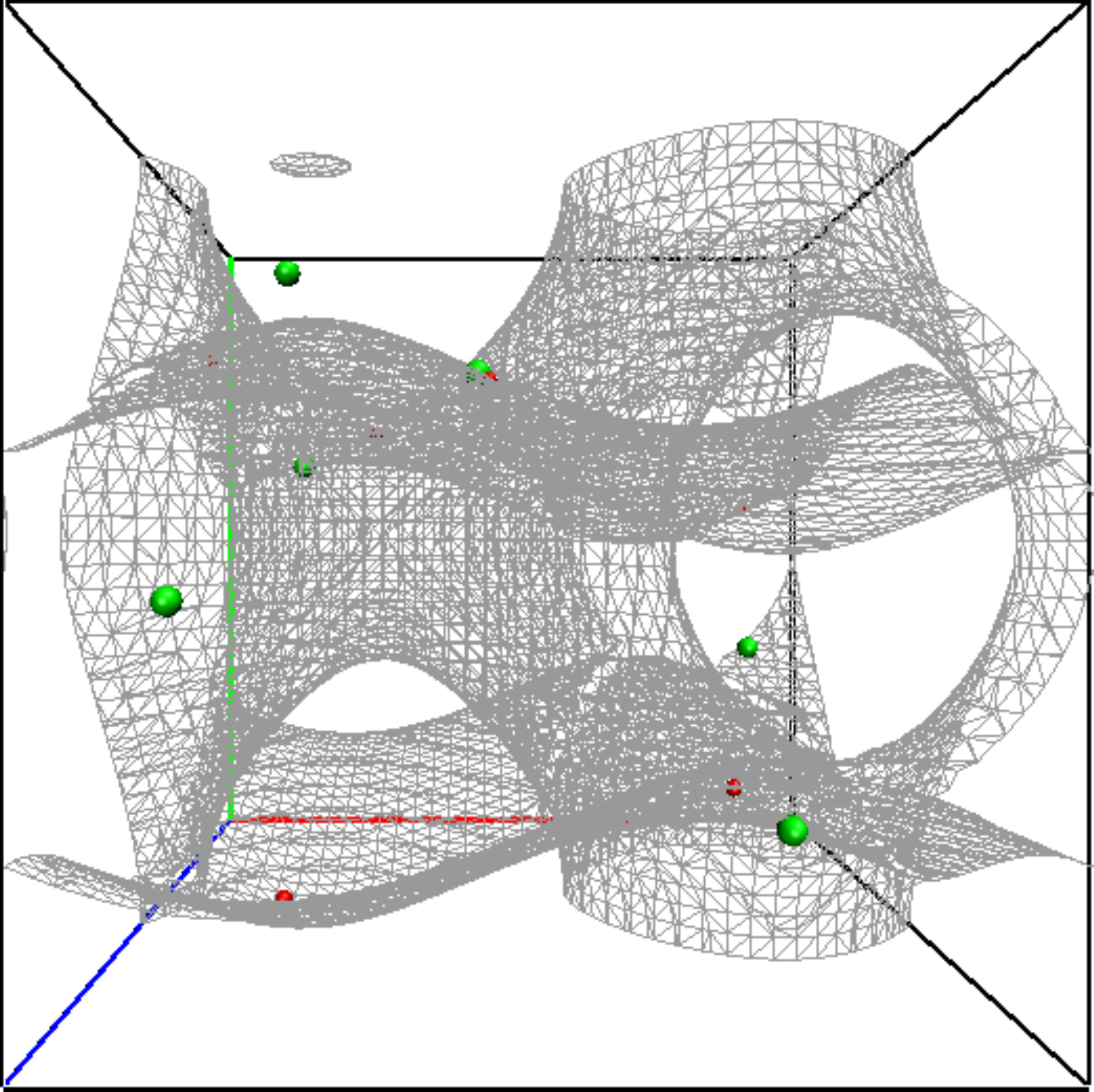}}}
     {\resizebox{1.1in}{!}{\includegraphics{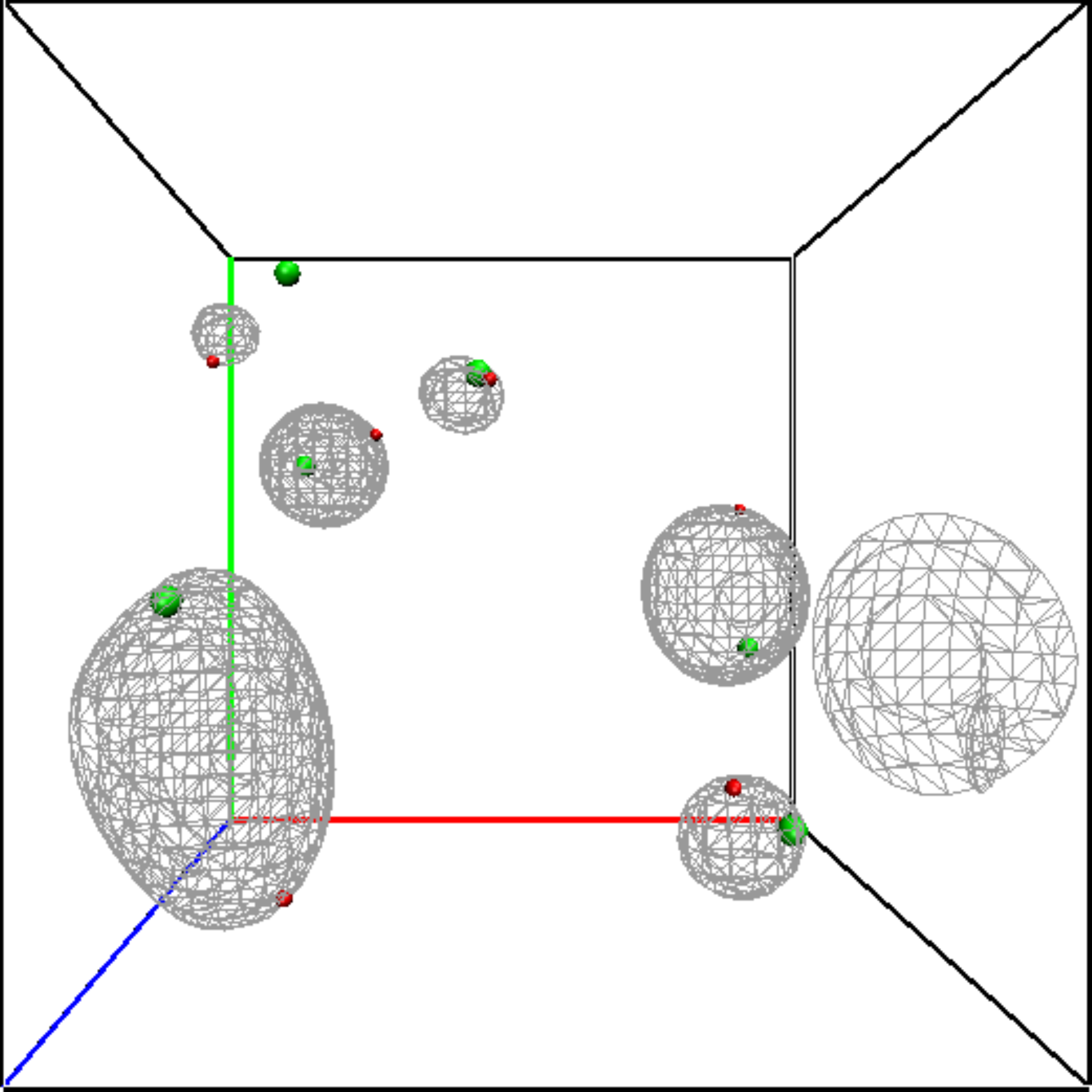}}} 
     {\resizebox{1.1in}{!}{\includegraphics{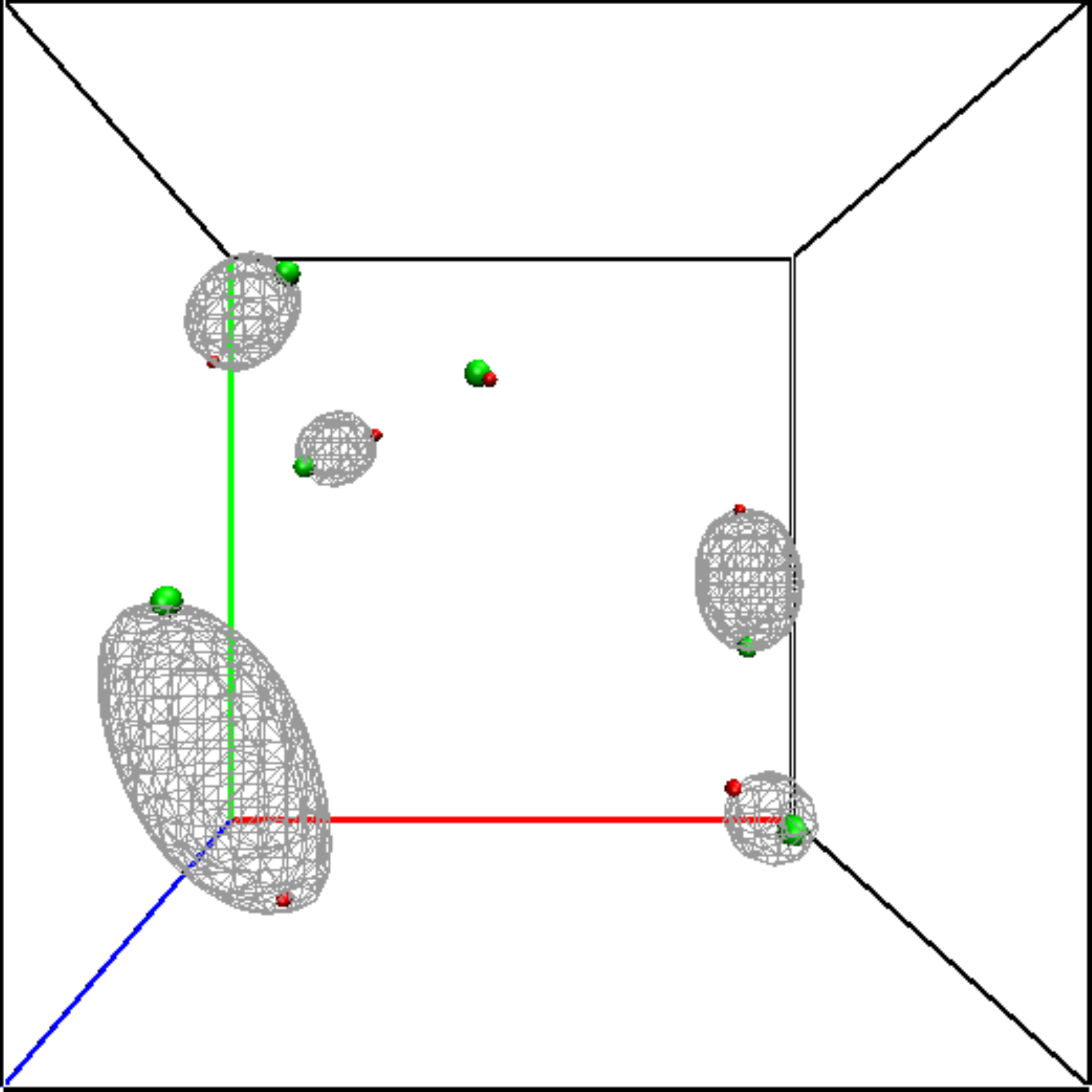}}}
    }
  \mbox{
     {\resizebox{1.1in}{!}{\includegraphics{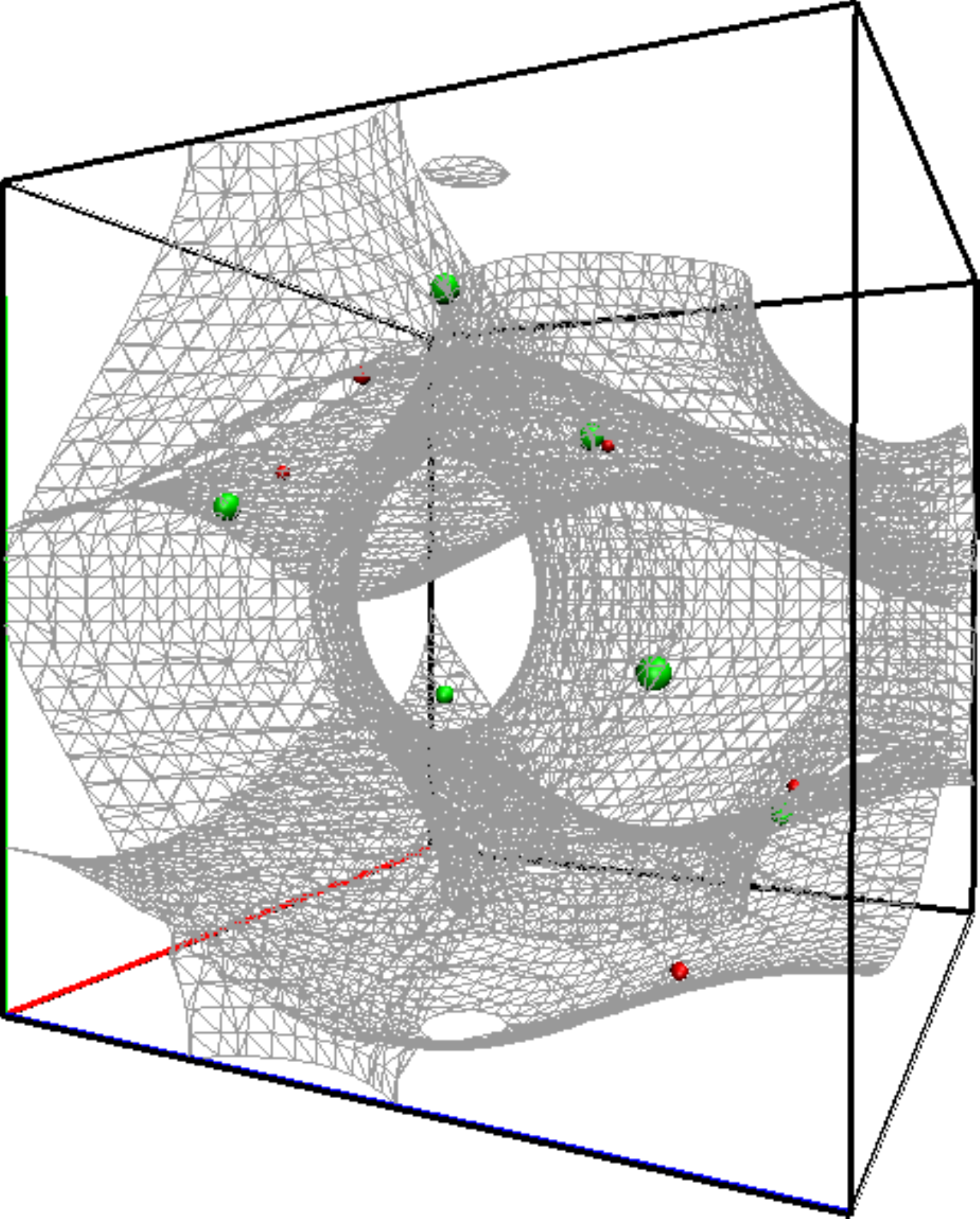}}}
     {\resizebox{1.1in}{!}{\includegraphics{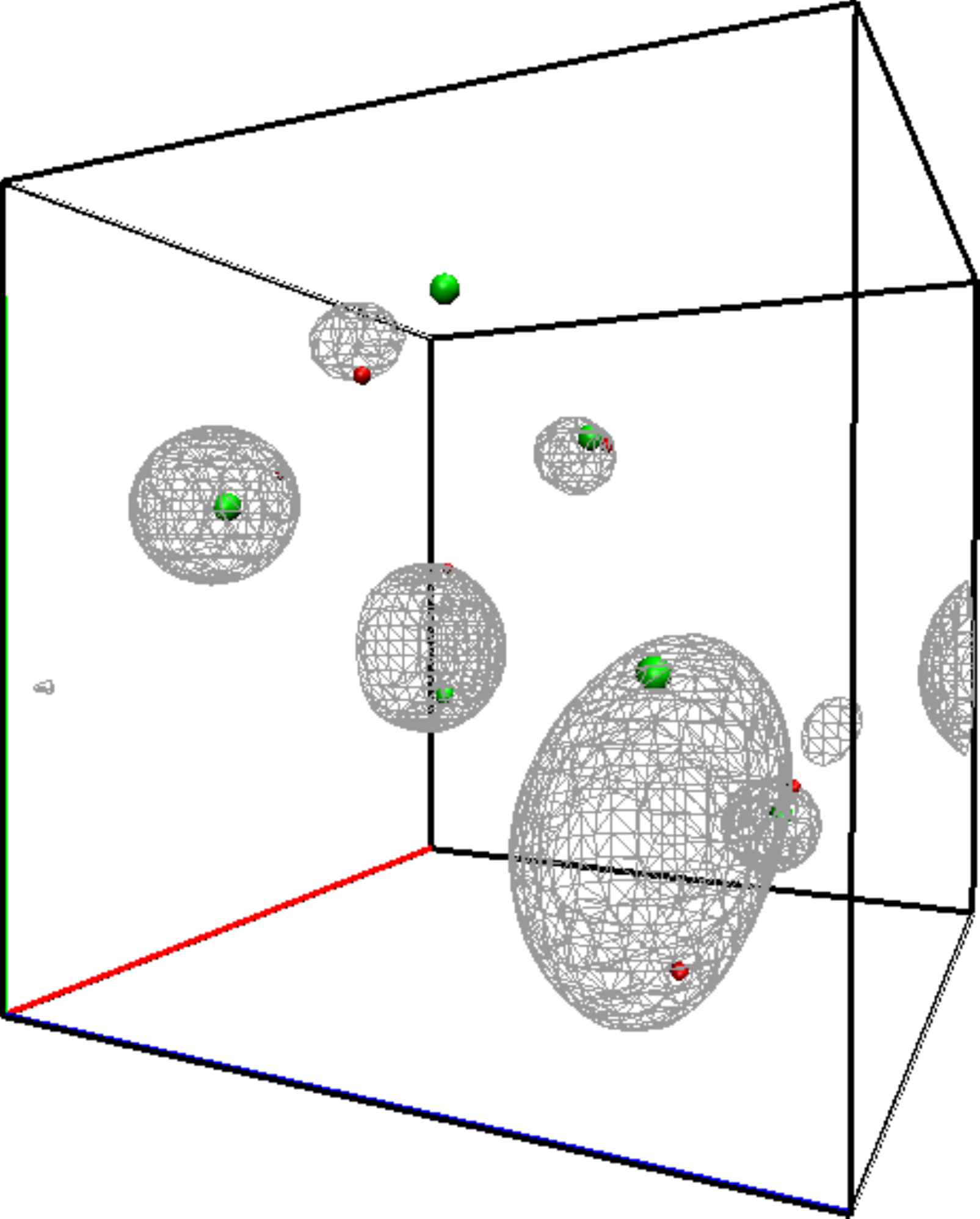}}} 
     {\resizebox{1.1in}{!}{\includegraphics{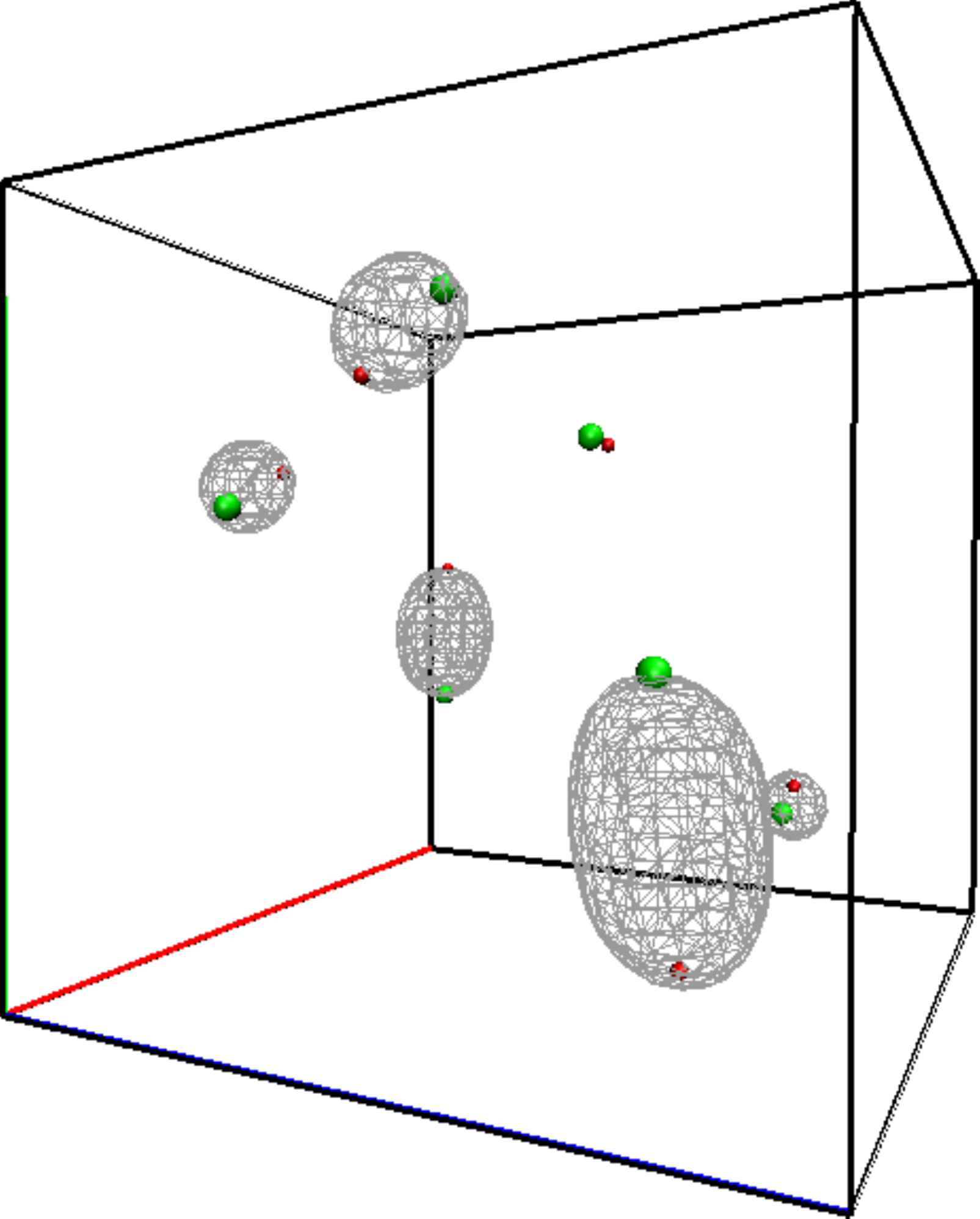}}}
   }
  \caption{Three-dimensional subsets of the nodal hypersurfaces for three types
    of wave functions and corresponding phases in the 14-atom system.
The node is obtained by scanning the simulation cell with a pair of
spin-up and spin-down atoms sitting on the top of each other while keeping the rest
of the atoms at fixed positions (tiny spheres). From the left to the right,
the columns show the nodal surfaces of the wave functions  corresponding
to the free Fermi gas,
the unitary limit and the BEC side of the crossover.
The lower row displays the same surfaces rotated by 45 degrees around the $z$-axis.
}
\label{node}
\end{figure}

To illustrate the character of the nodal surfaces in
the BCS-BEC systems, we present three-dimensional scans of the nodes for three wave functions corresponding
to the following scattering regimes:
the free atomic gas with no pairing, our best unitary-limit wave function, and the wave
function with enhanced pairing from the BEC side of the BEC-BCS phase
diagram ($a_sk_F=0.6592$).
The left column of Fig.~\ref{node} displays the nodal surface of the free atomic Fermi gas.
The delocalized nature of the system is apparent.
At the unitary limit, shown in the middle column of Fig.~\ref{node},
the shape of the nodal surface is significantly different as the pairing effects
clearly dominate and lead to a localized character of the nodes
from the perspective of a pair of up and down spin atoms.
The nodes on the BEC side (the right column) do not differ much from the unitary
limit, except for a slightly more pronounced localization.


\section{Conclusions}
We have carried
out QMC calculations of the zero-temperature, spin-unpolarized atomic Fermi gas
in the unitary limit.
We show that the interaction range impacts the resulting total energies significantly.
By extrapolating the interaction
range to zero we obtain the ratio $E/E_{\rm free}$
for 4, 14, 38 and 66 atoms to be 0.212(2), 0.407(2), 0.409(3) and 0.398(3), respectively.
Our results agree well with the previous fixed-node DMC
calculations when we employ similar simulation parameters such
as the atom density, the interaction range and the number of atoms.
From the released-node  DMC calculations for 4 and 14 atoms
we have found that the convergence
to the correct and asymptotically exact ground-state energies is unfavorably slow compared to the
growth of the statistical noise.
We were able to identify only
small energy gains within the simulation times that allowed for
acceptable signal to noise ratio.
We have calculated the two-body density matrix and the condensate
fraction in the limit of zero interaction range,
and we have found only small changes in these quantities when compared with
the previous calculations. Our condensate
fraction from the fixed-node DMC simulations is 0.56(1).

During the preparation of our manuscript we became aware of a similar
study where an interaction-range extrapolation was also
performed\cite{Gandolfi10}. The final result for the 66-atom system was
$\xi_{33,33}=0.383(2)$, which is approximately 4\% lower than ours. We
believe that a large portion of this difference can be attributed to differences
in the functional forms of the pairing orbital. Some influence could come also due to the
differences between the extrapolation methods employed in this work and in
Ref.~\cite{Gandolfi10}.


\begin{acknowledgments}
This work is supported by ARO and by the NSF grants
DMR-0804549 and OCI-0904794.
\end{acknowledgments}

\bibliographystyle{apsrev4-1}
\bibliography{unitaryfermigas}

\begin{thebibliography}{34}%
\makeatletter
\providecommand \@ifxundefined [1]{%
 \@ifx{#1\undefined}
}%
\providecommand \@ifnum [1]{%
 \ifnum #1\expandafter \@firstoftwo
 \else \expandafter \@secondoftwo
 \fi
}%
\providecommand \@ifx [1]{%
 \ifx #1\expandafter \@firstoftwo
 \else \expandafter \@secondoftwo
 \fi
}%
\providecommand \natexlab [1]{#1}%
\providecommand \enquote  [1]{``#1''}%
\providecommand \bibnamefont  [1]{#1}%
\providecommand \bibfnamefont [1]{#1}%
\providecommand \citenamefont [1]{#1}%
\providecommand \href@noop [0]{\@secondoftwo}%
\providecommand \href [0]{\begingroup \@sanitize@url \@href}%
\providecommand \@href[1]{\@@startlink{#1}\@@href}%
\providecommand \@@href[1]{\endgroup#1\@@endlink}%
\providecommand \@sanitize@url [0]{\catcode `\\12\catcode `\$12\catcode
  `\&12\catcode `\#12\catcode `\^12\catcode `\_12\catcode `\%12\relax}%
\providecommand \@@startlink[1]{}%
\providecommand \@@endlink[0]{}%
\providecommand \url  [0]{\begingroup\@sanitize@url \@url }%
\providecommand \@url [1]{\endgroup\@href {#1}{\urlprefix }}%
\providecommand \urlprefix  [0]{URL }%
\providecommand \Eprint [0]{\href }%
\providecommand \doibase [0]{http://dx.doi.org/}%
\providecommand \selectlanguage [0]{\@gobble}%
\providecommand \bibinfo  [0]{\@secondoftwo}%
\providecommand \bibfield  [0]{\@secondoftwo}%
\providecommand \translation [1]{[#1]}%
\providecommand \BibitemOpen [0]{}%
\providecommand \bibitemStop [0]{}%
\providecommand \bibitemNoStop [0]{.\EOS\space}%
\providecommand \EOS [0]{\spacefactor3000\relax}%
\providecommand \BibitemShut  [1]{\csname bibitem#1\endcsname}%
\let\auto@bib@innerbib\@empty
\bibitem [{\citenamefont {Ketterle}(2002)}]{Ketterle02}%
  \BibitemOpen
  \bibfield  {author} {\bibinfo {author} {\bibfnamefont {W.}~\bibnamefont
  {Ketterle}},\ }\href@noop {} {\bibfield  {journal} {\bibinfo  {journal}
  {Rev.\ Mod.\ Phys.}\ }\textbf {\bibinfo {volume} {74}},\ \bibinfo {pages}
  {1131} (\bibinfo {year} {2002})}\BibitemShut {NoStop}%
\bibitem [{\citenamefont {Ketterle}\ and\ \citenamefont
  {Zwierlein}(2008)}]{Ketterle08}%
  \BibitemOpen
  \bibfield  {author} {\bibinfo {author} {\bibfnamefont {W.}~\bibnamefont
  {Ketterle}}\ and\ \bibinfo {author} {\bibfnamefont {M.~W.}\ \bibnamefont
  {Zwierlein}},\ }\href@noop {} {\bibfield  {journal} {\bibinfo  {journal}
  {arXiv:0901.2500v1}\ } (\bibinfo {year} {2008})}\BibitemShut {NoStop}%
\bibitem [{\citenamefont {Giorgini}\ \emph {et~al.}(2008)\citenamefont
  {Giorgini}, \citenamefont {Pitaevskii},\ and\ \citenamefont
  {Stringari}}]{Giorgini08}%
  \BibitemOpen
  \bibfield  {author} {\bibinfo {author} {\bibfnamefont {S.}~\bibnamefont
  {Giorgini}}, \bibinfo {author} {\bibfnamefont {L.~P.}\ \bibnamefont
  {Pitaevskii}}, \ and\ \bibinfo {author} {\bibfnamefont {S.}~\bibnamefont
  {Stringari}},\ }\href@noop {} {\bibfield  {journal} {\bibinfo  {journal}
  {Rev.\ Mod.\ Phys.}\ }\textbf {\bibinfo {volume} {80}},\ \bibinfo {pages}
  {1215} (\bibinfo {year} {2008})}\BibitemShut {NoStop}%
\bibitem [{\citenamefont {Fano}(1961)}]{Fano61}%
  \BibitemOpen
  \bibfield  {author} {\bibinfo {author} {\bibfnamefont {U.}~\bibnamefont
  {Fano}},\ }\href@noop {} {\bibfield  {journal} {\bibinfo  {journal} {Phys.\
  Rev.}\ }\textbf {\bibinfo {volume} {124}},\ \bibinfo {pages} {1866} (\bibinfo
  {year} {1961})}\BibitemShut {NoStop}%
\bibitem [{\citenamefont {Feshbach}(1962)}]{Feshbach62}%
  \BibitemOpen
  \bibfield  {author} {\bibinfo {author} {\bibfnamefont {H.}~\bibnamefont
  {Feshbach}},\ }\href@noop {} {\bibfield  {journal} {\bibinfo  {journal}
  {Ann.\ Phys.\ (NY)}\ }\textbf {\bibinfo {volume} {19}},\ \bibinfo {pages}
  {287} (\bibinfo {year} {1962})}\BibitemShut {NoStop}%
\bibitem [{\citenamefont {Inouye}\ \emph {et~al.}(1998)\citenamefont {Inouye},
  \citenamefont {Andrews}, \citenamefont {Stenger}, \citenamefont {Miesner},
  \citenamefont {Stamper-Kurn},\ and\ \citenamefont {Ketterle}}]{Ketterle98}%
  \BibitemOpen
  \bibfield  {author} {\bibinfo {author} {\bibfnamefont {S.}~\bibnamefont
  {Inouye}}, \bibinfo {author} {\bibfnamefont {M.~R.}\ \bibnamefont {Andrews}},
  \bibinfo {author} {\bibfnamefont {J.}~\bibnamefont {Stenger}}, \bibinfo
  {author} {\bibfnamefont {H.~J.}\ \bibnamefont {Miesner}}, \bibinfo {author}
  {\bibfnamefont {D.~M.}\ \bibnamefont {Stamper-Kurn}}, \ and\ \bibinfo
  {author} {\bibfnamefont {W.}~\bibnamefont {Ketterle}},\ }\href@noop {}
  {\bibfield  {journal} {\bibinfo  {journal} {Nature}\ }\textbf {\bibinfo
  {volume} {392}},\ \bibinfo {pages} {151} (\bibinfo {year}
  {1998})}\BibitemShut {NoStop}%
\bibitem [{\citenamefont {Courteille}\ \emph {et~al.}(1998)\citenamefont
  {Courteille}, \citenamefont {Freeland}, \citenamefont {Heinzen},
  \citenamefont {van Abeelen},\ and\ \citenamefont {Verhaar}}]{Verhaar98}%
  \BibitemOpen
  \bibfield  {author} {\bibinfo {author} {\bibfnamefont {P.}~\bibnamefont
  {Courteille}}, \bibinfo {author} {\bibfnamefont {R.~S.}\ \bibnamefont
  {Freeland}}, \bibinfo {author} {\bibfnamefont {D.~J.}\ \bibnamefont
  {Heinzen}}, \bibinfo {author} {\bibfnamefont {F.~A.}\ \bibnamefont {van
  Abeelen}}, \ and\ \bibinfo {author} {\bibfnamefont {B.~J.}\ \bibnamefont
  {Verhaar}},\ }\href@noop {} {\bibfield  {journal} {\bibinfo  {journal}
  {Phys.\ Rev.\ Lett.}\ }\textbf {\bibinfo {volume} {81}},\ \bibinfo {pages}
  {69} (\bibinfo {year} {1998})}\BibitemShut {NoStop}%
\bibitem [{\citenamefont {Bartenstein}\ \emph {et~al.}(2004)\citenamefont
  {Bartenstein}, \citenamefont {Altmeyer}, \citenamefont {Riedl}, \citenamefont
  {Jochim}, \citenamefont {Chin}, \citenamefont {Denschlag},\ and\
  \citenamefont {Grimm}}]{Bartenstein04}%
  \BibitemOpen
  \bibfield  {author} {\bibinfo {author} {\bibfnamefont {M.}~\bibnamefont
  {Bartenstein}}, \bibinfo {author} {\bibfnamefont {A.}~\bibnamefont
  {Altmeyer}}, \bibinfo {author} {\bibfnamefont {S.}~\bibnamefont {Riedl}},
  \bibinfo {author} {\bibfnamefont {S.}~\bibnamefont {Jochim}}, \bibinfo
  {author} {\bibfnamefont {C.}~\bibnamefont {Chin}}, \bibinfo {author}
  {\bibfnamefont {J.~H.}\ \bibnamefont {Denschlag}}, \ and\ \bibinfo {author}
  {\bibfnamefont {R.}~\bibnamefont {Grimm}},\ }\href@noop {} {\bibfield
  {journal} {\bibinfo  {journal} {Phys.\ Rev.\ Lett.}\ }\textbf {\bibinfo
  {volume} {92}},\ \bibinfo {pages} {120401} (\bibinfo {year}
  {2004})}\BibitemShut {NoStop}%
\bibitem [{\citenamefont {Bourdel}\ \emph {et~al.}(2004)\citenamefont
  {Bourdel}, \citenamefont {Khaykovich}, \citenamefont {Cubizolles},
  \citenamefont {Zhang}, \citenamefont {Chevy}, \citenamefont {Teichmann},
  \citenamefont {Tarruell}, \citenamefont {Kokkelmans},\ and\ \citenamefont
  {Salomon}}]{Bourdel04}%
  \BibitemOpen
  \bibfield  {author} {\bibinfo {author} {\bibfnamefont {T.}~\bibnamefont
  {Bourdel}}, \bibinfo {author} {\bibfnamefont {L.}~\bibnamefont {Khaykovich}},
  \bibinfo {author} {\bibfnamefont {J.}~\bibnamefont {Cubizolles}}, \bibinfo
  {author} {\bibfnamefont {J.}~\bibnamefont {Zhang}}, \bibinfo {author}
  {\bibfnamefont {F.}~\bibnamefont {Chevy}}, \bibinfo {author} {\bibfnamefont
  {M.}~\bibnamefont {Teichmann}}, \bibinfo {author} {\bibfnamefont
  {L.}~\bibnamefont {Tarruell}}, \bibinfo {author} {\bibfnamefont {S.~J. J.
  M.~F.}\ \bibnamefont {Kokkelmans}}, \ and\ \bibinfo {author} {\bibfnamefont
  {C.}~\bibnamefont {Salomon}},\ }\href@noop {} {\bibfield  {journal} {\bibinfo
   {journal} {Phys.\ Rev.\ Lett.}\ }\textbf {\bibinfo {volume} {93}},\ \bibinfo
  {pages} {050401} (\bibinfo {year} {2004})}\BibitemShut {NoStop}%
\bibitem [{\citenamefont {Kinast}\ \emph {et~al.}(2005)\citenamefont {Kinast},
  \citenamefont {Turlapov}, \citenamefont {Thomas}, \citenamefont {Chen},
  \citenamefont {Stajic},\ and\ \citenamefont {Levin}}]{Thomas05}%
  \BibitemOpen
  \bibfield  {author} {\bibinfo {author} {\bibfnamefont {J.}~\bibnamefont
  {Kinast}}, \bibinfo {author} {\bibfnamefont {A.}~\bibnamefont {Turlapov}},
  \bibinfo {author} {\bibfnamefont {J.~E.}\ \bibnamefont {Thomas}}, \bibinfo
  {author} {\bibfnamefont {Q.}~\bibnamefont {Chen}}, \bibinfo {author}
  {\bibfnamefont {J.}~\bibnamefont {Stajic}}, \ and\ \bibinfo {author}
  {\bibfnamefont {K.}~\bibnamefont {Levin}},\ }\href@noop {} {\bibfield
  {journal} {\bibinfo  {journal} {Science}\ }\textbf {\bibinfo {volume}
  {307}},\ \bibinfo {pages} {1296} (\bibinfo {year} {2005})}\BibitemShut
  {NoStop}%
\bibitem [{\citenamefont {Stewart}\ \emph {et~al.}(2006)\citenamefont
  {Stewart}, \citenamefont {Gaebler}, \citenamefont {Regal},\ and\
  \citenamefont {Jin}}]{Stewart06}%
  \BibitemOpen
  \bibfield  {author} {\bibinfo {author} {\bibfnamefont {J.~T.}\ \bibnamefont
  {Stewart}}, \bibinfo {author} {\bibfnamefont {J.~P.}\ \bibnamefont
  {Gaebler}}, \bibinfo {author} {\bibfnamefont {C.~A.}\ \bibnamefont {Regal}},
  \ and\ \bibinfo {author} {\bibfnamefont {D.~S.}\ \bibnamefont {Jin}},\
  }\href@noop {} {\bibfield  {journal} {\bibinfo  {journal} {Phys.\ Rev.\
  Lett.}\ }\textbf {\bibinfo {volume} {97}},\ \bibinfo {pages} {220406}
  (\bibinfo {year} {2006})}\BibitemShut {NoStop}%
\bibitem [{\citenamefont {Joseph}\ \emph {et~al.}(2007)\citenamefont {Joseph},
  \citenamefont {Clancy}, \citenamefont {Luo}, \citenamefont {Kinast},
  \citenamefont {Turlapov},\ and\ \citenamefont {Thomas}}]{Thomas07}%
  \BibitemOpen
  \bibfield  {author} {\bibinfo {author} {\bibfnamefont {J.}~\bibnamefont
  {Joseph}}, \bibinfo {author} {\bibfnamefont {B.}~\bibnamefont {Clancy}},
  \bibinfo {author} {\bibfnamefont {L.}~\bibnamefont {Luo}}, \bibinfo {author}
  {\bibfnamefont {J.}~\bibnamefont {Kinast}}, \bibinfo {author} {\bibfnamefont
  {A.}~\bibnamefont {Turlapov}}, \ and\ \bibinfo {author} {\bibfnamefont
  {J.~E.}\ \bibnamefont {Thomas}},\ }\href@noop {} {\bibfield  {journal}
  {\bibinfo  {journal} {Phys.\ Rev.\ Lett.}\ }\textbf {\bibinfo {volume}
  {98}},\ \bibinfo {pages} {170401} (\bibinfo {year} {2007})}\BibitemShut
  {NoStop}%
\bibitem [{\citenamefont {Carlson}\ \emph {et~al.}(2003)\citenamefont
  {Carlson}, \citenamefont {Chang}, \citenamefont {Pandharipande},\ and\
  \citenamefont {Schmidt}}]{Carlson03}%
  \BibitemOpen
  \bibfield  {author} {\bibinfo {author} {\bibfnamefont {J.}~\bibnamefont
  {Carlson}}, \bibinfo {author} {\bibfnamefont {S.~Y.}\ \bibnamefont {Chang}},
  \bibinfo {author} {\bibfnamefont {V.~R.}\ \bibnamefont {Pandharipande}}, \
  and\ \bibinfo {author} {\bibfnamefont {K.~E.}\ \bibnamefont {Schmidt}},\
  }\href@noop {} {\bibfield  {journal} {\bibinfo  {journal} {Phys.\ Rev.\
  Lett.}\ }\textbf {\bibinfo {volume} {91}},\ \bibinfo {pages} {050401}
  (\bibinfo {year} {2003})}\BibitemShut {NoStop}%
\bibitem [{\citenamefont {Chang}\ \emph {et~al.}(2004)\citenamefont {Chang},
  \citenamefont {Pandharipande}, \citenamefont {Carlson},\ and\ \citenamefont
  {Schmidt}}]{Carlson04}%
  \BibitemOpen
  \bibfield  {author} {\bibinfo {author} {\bibfnamefont {S.~Y.}\ \bibnamefont
  {Chang}}, \bibinfo {author} {\bibfnamefont {V.~R.}\ \bibnamefont
  {Pandharipande}}, \bibinfo {author} {\bibfnamefont {J.}~\bibnamefont
  {Carlson}}, \ and\ \bibinfo {author} {\bibfnamefont {K.~E.}\ \bibnamefont
  {Schmidt}},\ }\href@noop {} {\bibfield  {journal} {\bibinfo  {journal}
  {Phys.\ Rev.\ A}\ }\textbf {\bibinfo {volume} {70}},\ \bibinfo {pages}
  {043602} (\bibinfo {year} {2004})}\BibitemShut {NoStop}%
\bibitem [{\citenamefont {Astrakharchik}\ \emph {et~al.}(2004)\citenamefont
  {Astrakharchik}, \citenamefont {Boronat}, \citenamefont {Casulleras},\ and\
  \citenamefont {Giorgini}}]{Astrak04}%
  \BibitemOpen
  \bibfield  {author} {\bibinfo {author} {\bibfnamefont {G.~E.}\ \bibnamefont
  {Astrakharchik}}, \bibinfo {author} {\bibfnamefont {J.}~\bibnamefont
  {Boronat}}, \bibinfo {author} {\bibfnamefont {J.}~\bibnamefont {Casulleras}},
  \ and\ \bibinfo {author} {\bibfnamefont {S.}~\bibnamefont {Giorgini}},\
  }\href@noop {} {\bibfield  {journal} {\bibinfo  {journal} {Phys.\ Rev.\
  Lett.}\ }\textbf {\bibinfo {volume} {93}},\ \bibinfo {pages} {200404}
  (\bibinfo {year} {2004})}\BibitemShut {NoStop}%
\bibitem [{\citenamefont {Astrakharchik}\ \emph {et~al.}(2005)\citenamefont
  {Astrakharchik}, \citenamefont {Boronat}, \citenamefont {Casulleras},\ and\
  \citenamefont {Giorgini}}]{Astrak05}%
  \BibitemOpen
  \bibfield  {author} {\bibinfo {author} {\bibfnamefont {G.~E.}\ \bibnamefont
  {Astrakharchik}}, \bibinfo {author} {\bibfnamefont {J.}~\bibnamefont
  {Boronat}}, \bibinfo {author} {\bibfnamefont {J.}~\bibnamefont {Casulleras}},
  \ and\ \bibinfo {author} {\bibfnamefont {S.}~\bibnamefont {Giorgini}},\
  }\href@noop {} {\bibfield  {journal} {\bibinfo  {journal} {Phys.\ Rev.\
  Lett.}\ }\textbf {\bibinfo {volume} {95}},\ \bibinfo {pages} {230405}
  (\bibinfo {year} {2005})}\BibitemShut {NoStop}%
\bibitem [{\citenamefont {Chang}\ and\ \citenamefont
  {Pandharipande}(2005)}]{Pandharipande05}%
  \BibitemOpen
  \bibfield  {author} {\bibinfo {author} {\bibfnamefont {S.~Y.}\ \bibnamefont
  {Chang}}\ and\ \bibinfo {author} {\bibfnamefont {V.~R.}\ \bibnamefont
  {Pandharipande}},\ }\href@noop {} {\bibfield  {journal} {\bibinfo  {journal}
  {Phys.\ Rev.\ Lett.}\ }\textbf {\bibinfo {volume} {95}},\ \bibinfo {pages}
  {080402} (\bibinfo {year} {2005})}\BibitemShut {NoStop}%
\bibitem [{\citenamefont {Morris}\ \emph {et~al.}(2010)\citenamefont {Morris},
  \citenamefont {L\'opez~R\'\i{}os},\ and\ \citenamefont {Needs}}]{Needs10}%
  \BibitemOpen
  \bibfield  {author} {\bibinfo {author} {\bibfnamefont {A.~J.}\ \bibnamefont
  {Morris}}, \bibinfo {author} {\bibfnamefont {P.}~\bibnamefont
  {L\'opez~R\'\i{}os}}, \ and\ \bibinfo {author} {\bibfnamefont {R.~J.}\
  \bibnamefont {Needs}},\ }\href@noop {} {\bibfield  {journal} {\bibinfo
  {journal} {Phys.\ Rev.\ A}\ }\textbf {\bibinfo {volume} {81}},\ \bibinfo
  {pages} {033619} (\bibinfo {year} {2010})}\BibitemShut {NoStop}%
\bibitem [{\citenamefont {Lee}(2006)}]{Lee06}%
  \BibitemOpen
  \bibfield  {author} {\bibinfo {author} {\bibfnamefont {D.}~\bibnamefont
  {Lee}},\ }\href@noop {} {\bibfield  {journal} {\bibinfo  {journal} {Phys.\
  Rev.\ B}\ }\textbf {\bibinfo {volume} {73}},\ \bibinfo {pages} {115112}
  (\bibinfo {year} {2006})}\BibitemShut {NoStop}%
\bibitem [{\citenamefont {Lee}(2008)}]{Lee08}%
  \BibitemOpen
  \bibfield  {author} {\bibinfo {author} {\bibfnamefont {D.}~\bibnamefont
  {Lee}},\ }\href@noop {} {\bibfield  {journal} {\bibinfo  {journal} {Phys.
  Rev. C}\ }\textbf {\bibinfo {volume} {78}},\ \bibinfo {pages} {024001}
  (\bibinfo {year} {2008})}\BibitemShut {NoStop}%
\bibitem [{\citenamefont {Akkineni}\ \emph {et~al.}(2007)\citenamefont
  {Akkineni}, \citenamefont {Ceperley},\ and\ \citenamefont
  {Trivedi}}]{Trivedi07}%
  \BibitemOpen
  \bibfield  {author} {\bibinfo {author} {\bibfnamefont {V.~K.}\ \bibnamefont
  {Akkineni}}, \bibinfo {author} {\bibfnamefont {D.~M.}\ \bibnamefont
  {Ceperley}}, \ and\ \bibinfo {author} {\bibfnamefont {N.}~\bibnamefont
  {Trivedi}},\ }\href@noop {} {\bibfield  {journal} {\bibinfo  {journal}
  {Phys.\ Rev.\ B}\ }\textbf {\bibinfo {volume} {76}},\ \bibinfo {pages}
  {165116} (\bibinfo {year} {2007})}\BibitemShut {NoStop}%
\bibitem [{\citenamefont {Burovski}\ \emph {et~al.}(2006)\citenamefont
  {Burovski}, \citenamefont {Prokof'ev}, \citenamefont {Svistunov},\ and\
  \citenamefont {Troyer}}]{Burovski06}%
  \BibitemOpen
  \bibfield  {author} {\bibinfo {author} {\bibfnamefont {E.}~\bibnamefont
  {Burovski}}, \bibinfo {author} {\bibfnamefont {N.}~\bibnamefont {Prokof'ev}},
  \bibinfo {author} {\bibfnamefont {B.}~\bibnamefont {Svistunov}}, \ and\
  \bibinfo {author} {\bibfnamefont {M.}~\bibnamefont {Troyer}},\ }\href@noop {}
  {\bibfield  {journal} {\bibinfo  {journal} {Phys.\ Rev.\ Lett.}\ }\textbf
  {\bibinfo {volume} {96}},\ \bibinfo {pages} {160402} (\bibinfo {year}
  {2006})}\BibitemShut {NoStop}%
\bibitem [{\citenamefont {Gurarie}\ and\ \citenamefont
  {Radzihovsky}(2007)}]{Gurarie07}%
  \BibitemOpen
  \bibfield  {author} {\bibinfo {author} {\bibfnamefont {V.}~\bibnamefont
  {Gurarie}}\ and\ \bibinfo {author} {\bibfnamefont {L.}~\bibnamefont
  {Radzihovsky}},\ }\href@noop {} {\bibfield  {journal} {\bibinfo  {journal}
  {Ann.\ Phys.}\ }\textbf {\bibinfo {volume} {322}},\ \bibinfo {pages} {2}
  (\bibinfo {year} {2007})}\BibitemShut {NoStop}%
\bibitem [{\citenamefont {Bulgac}\ \emph {et~al.}(2006)\citenamefont {Bulgac},
  \citenamefont {Drut},\ and\ \citenamefont {Magierski}}]{Bulgac06}%
  \BibitemOpen
  \bibfield  {author} {\bibinfo {author} {\bibfnamefont {A.}~\bibnamefont
  {Bulgac}}, \bibinfo {author} {\bibfnamefont {J.~E.}\ \bibnamefont {Drut}}, \
  and\ \bibinfo {author} {\bibfnamefont {P.}~\bibnamefont {Magierski}},\
  }\href@noop {} {\bibfield  {journal} {\bibinfo  {journal} {Phys.\ Rev.\
  Lett.}\ }\textbf {\bibinfo {volume} {96}},\ \bibinfo {pages} {090404}
  (\bibinfo {year} {2006})}\BibitemShut {NoStop}%
\bibitem [{\citenamefont {Bulgac}\ \emph {et~al.}(2008)\citenamefont {Bulgac},
  \citenamefont {Drut},\ and\ \citenamefont {Magierski}}]{Bulgac08}%
  \BibitemOpen
  \bibfield  {author} {\bibinfo {author} {\bibfnamefont {A.}~\bibnamefont
  {Bulgac}}, \bibinfo {author} {\bibfnamefont {J.~E.}\ \bibnamefont {Drut}}, \
  and\ \bibinfo {author} {\bibfnamefont {P.}~\bibnamefont {Magierski}},\
  }\href@noop {} {\bibfield  {journal} {\bibinfo  {journal} {Phys.\ Rev.\ A}\
  }\textbf {\bibinfo {volume} {78}},\ \bibinfo {pages} {023625} (\bibinfo
  {year} {2008})}\BibitemShut {NoStop}%
\bibitem [{\citenamefont {Abe}\ and\ \citenamefont
  {Seki}(2009{\natexlab{a}})}]{Abe09a}%
  \BibitemOpen
  \bibfield  {author} {\bibinfo {author} {\bibfnamefont {T.}~\bibnamefont
  {Abe}}\ and\ \bibinfo {author} {\bibfnamefont {R.}~\bibnamefont {Seki}},\
  }\href@noop {} {\bibfield  {journal} {\bibinfo  {journal} {Phys.\ Rev.\ C}\
  }\textbf {\bibinfo {volume} {79}},\ \bibinfo {pages} {054002} (\bibinfo
  {year} {2009}{\natexlab{a}})}\BibitemShut {NoStop}%
\bibitem [{\citenamefont {Abe}\ and\ \citenamefont
  {Seki}(2009{\natexlab{b}})}]{Abe09b}%
  \BibitemOpen
  \bibfield  {author} {\bibinfo {author} {\bibfnamefont {T.}~\bibnamefont
  {Abe}}\ and\ \bibinfo {author} {\bibfnamefont {R.}~\bibnamefont {Seki}},\
  }\href@noop {} {\bibfield  {journal} {\bibinfo  {journal} {Phys.\ Rev.\ C}\
  }\textbf {\bibinfo {volume} {79}},\ \bibinfo {pages} {054003} (\bibinfo
  {year} {2009}{\natexlab{b}})}\BibitemShut {NoStop}%
\bibitem [{\citenamefont {Foulkes}\ \emph {et~al.}(2001)\citenamefont
  {Foulkes}, \citenamefont {Mitas}, \citenamefont {Needs},\ and\ \citenamefont
  {Rajagopal}}]{Mitas01}%
  \BibitemOpen
  \bibfield  {author} {\bibinfo {author} {\bibfnamefont {W.}~\bibnamefont
  {Foulkes}}, \bibinfo {author} {\bibfnamefont {L.}~\bibnamefont {Mitas}},
  \bibinfo {author} {\bibfnamefont {R.~J.}\ \bibnamefont {Needs}}, \ and\
  \bibinfo {author} {\bibfnamefont {G.}~\bibnamefont {Rajagopal}},\ }\href@noop
  {} {\bibfield  {journal} {\bibinfo  {journal} {Rev. Mod. Phys.}\ }\textbf
  {\bibinfo {volume} {73}},\ \bibinfo {pages} {33} (\bibinfo {year}
  {2001})}\BibitemShut {NoStop}%
\bibitem [{\citenamefont {Ceperley}\ and\ \citenamefont
  {Alder}(1980)}]{Ceperley80}%
  \BibitemOpen
  \bibfield  {author} {\bibinfo {author} {\bibfnamefont {D.~M.}\ \bibnamefont
  {Ceperley}}\ and\ \bibinfo {author} {\bibfnamefont {B.~J.}\ \bibnamefont
  {Alder}},\ }\href@noop {} {\bibfield  {journal} {\bibinfo  {journal} {Phys.\
  Rev.\ Lett.}\ }\textbf {\bibinfo {volume} {45}},\ \bibinfo {pages} {566}
  (\bibinfo {year} {1980})}\BibitemShut {NoStop}%
\bibitem [{\citenamefont {Ceperley}\ and\ \citenamefont
  {Alder}(1984)}]{Ceperley84}%
  \BibitemOpen
  \bibfield  {author} {\bibinfo {author} {\bibfnamefont {D.~M.}\ \bibnamefont
  {Ceperley}}\ and\ \bibinfo {author} {\bibfnamefont {B.~J.}\ \bibnamefont
  {Alder}},\ }\href@noop {} {\bibfield  {journal} {\bibinfo  {journal} {J.\
  Chem.\ Phys.}\ }\textbf {\bibinfo {volume} {81}},\ \bibinfo {pages} {5833}
  (\bibinfo {year} {1984})}\BibitemShut {NoStop}%
\bibitem [{\citenamefont {Umrigar}\ and\ \citenamefont
  {Filippi}(2005)}]{Umrigar05}%
  \BibitemOpen
  \bibfield  {author} {\bibinfo {author} {\bibfnamefont {C.~J.}\ \bibnamefont
  {Umrigar}}\ and\ \bibinfo {author} {\bibfnamefont {C.}~\bibnamefont
  {Filippi}},\ }\href@noop {} {\bibfield  {journal} {\bibinfo  {journal} {Phys.
  Rev. Lett.}\ }\textbf {\bibinfo {volume} {94}},\ \bibinfo {pages} {150201}
  (\bibinfo {year} {2005})}\BibitemShut {NoStop}%
\bibitem [{\citenamefont {Casulleras}\ and\ \citenamefont
  {Boronat}(2000)}]{Casulleras00}%
  \BibitemOpen
  \bibfield  {author} {\bibinfo {author} {\bibfnamefont {J.}~\bibnamefont
  {Casulleras}}\ and\ \bibinfo {author} {\bibfnamefont {J.}~\bibnamefont
  {Boronat}},\ }\href@noop {} {\bibfield  {journal} {\bibinfo  {journal}
  {Phys.\ Rev.\ Lett.}\ }\textbf {\bibinfo {volume} {84}},\ \bibinfo {pages}
  {3121} (\bibinfo {year} {2000})}\BibitemShut {NoStop}%
\bibitem [{\citenamefont {Bour}\ \emph {et~al.}(2011)\citenamefont {Bour},
  \citenamefont {Li}, \citenamefont {Lee}, \citenamefont {Mei\ss{}ner},\ and\
  \citenamefont {Mitas}}]{Lee11}%
  \BibitemOpen
  \bibfield  {author} {\bibinfo {author} {\bibfnamefont {S.}~\bibnamefont
  {Bour}}, \bibinfo {author} {\bibfnamefont {X.}~\bibnamefont {Li}}, \bibinfo
  {author} {\bibfnamefont {D.}~\bibnamefont {Lee}}, \bibinfo {author}
  {\bibfnamefont {U.-G.}\ \bibnamefont {Mei\ss{}ner}}, \ and\ \bibinfo {author}
  {\bibfnamefont {L.}~\bibnamefont {Mitas}},\ }\href@noop {} {\bibfield
  {journal} {\bibinfo  {journal} {arXiv:1104.2102}\ } (\bibinfo {year}
  {2011})}\BibitemShut {NoStop}%
\bibitem [{\citenamefont {Forbes}\ \emph {et~al.}(2010)\citenamefont {Forbes},
  \citenamefont {Gandolfi},\ and\ \citenamefont {Gezerlis}}]{Gandolfi10}%
  \BibitemOpen
  \bibfield  {author} {\bibinfo {author} {\bibfnamefont {M.~M.}\ \bibnamefont
  {Forbes}}, \bibinfo {author} {\bibfnamefont {S.}~\bibnamefont {Gandolfi}}, \
  and\ \bibinfo {author} {\bibfnamefont {A.}~\bibnamefont {Gezerlis}},\
  }\href@noop {} {\bibfield  {journal} {\bibinfo  {journal} {arXiv:1011.2197}\
  } (\bibinfo {year} {2010})}\BibitemShut {NoStop}%
\end{thebibliography}%

\end{document}